\documentclass[onecolumn,10pt,aps,showpacs,longbibliography,superscriptaddress]{revtex4-2}

\usepackage{graphicx}
\usepackage{amsmath}
\usepackage{amsfonts}
\usepackage{amssymb}
\usepackage{color}
\usepackage[dvipsnames]{xcolor}
\usepackage{float}
\usepackage{mathrsfs}
\usepackage{multirow}

\newcommand {\ra}{\rightarrow}

\newcommand\undermat[2]{%
  \makebox[0pt][l]{$\smash{\underbrace{\phantom{%
    \begin{matrix}#2\end{matrix}}}_{\text{$#1$}}}$}#2}

\begin{document}

\title{Numerical direct scattering transform for breathers}

\author{I. Mullyadzhanov}
\affiliation{Institute of Automation and Electrometry SB RAS, Novosibirsk 630090, Russia}
\affiliation{Skolkovo Institute of Science and Technology, Moscow, 121205, Russia}

\author{A. Gudko}
\affiliation{Institute of Thermophysics SB RAS, Novosibirsk 630090, Russia}
\affiliation{Novosibirsk State University, Novosibirsk 630090, Russia}

\author{R. Mullyadzhanov}
\affiliation{Institute of Thermophysics SB RAS, Novosibirsk 630090, Russia}
\affiliation{Novosibirsk State University, Novosibirsk 630090, Russia}

\author{A. Gelash}
\email[Corresponding author: ]{Andrey.Gelash@u-bourgogne.fr}
\affiliation{Laboratoire Interdisciplinaire Carnot de Bourgogne (ICB), UMR 6303 CNRS -- Université Bourgogne Franche-Comté, 21078 Dijon, France}

\begin{abstract}
We consider the model of the focusing one-dimensional nonlinear Schrödinger equation (fNLSE) in the presence of an unstable constant background, which exhibits coherent solitary wave structures -- breathers. 
Within the inverse scattering transform (IST) method, we study the problem of the scattering data numerical computation for a broad class of breathers localized in space.
Such direct scattering transform (DST) procedure requires a numerical solution of the auxiliary Zakharov--Shabat system with boundary conditions corresponding to the background.
To find the solution we compute the transfer matrix using the second-order Boffetta--Osborne approach and recently developed high-order numerical schemes based on the Magnus expansion.
To recover the scattering data of breathers, we derive analytical relations between the scattering coefficients and the transfer matrix elements.
Then we construct localized single- and multi-breather solutions and verify the developed numerical approach by accurately recovering the complete set of the scattering data, which provides us with information about the amplitude, velocity, phase, and position of each breather.
To combine the conventional IST approach with the efficient dressing method for multi-breather solutions, we provide the exact relation between the parameters of breathers in these two frameworks.
\end{abstract}

\maketitle

%-------------------------------------------------------------------------------------------------------------
%-------------------------------------------------------------------------------------------------------------

\section{Introduction}
\label{Sec:Intro}

Nonlinear dynamics of coherent solitary wave structures emerging in many physical systems ranging from optical fiber links and fluid reservoirs to Bose--Einstein condensates represent a wide field of experimental, theoretical, and numerical studies \cite{remoissenet2013waves, akhmediev1997book, newell1985solitons, pelinovsky2008book}. 
The fundamental mechanism of modulation instability makes the behavior of solitary waves intricate by generating a broad spectrum of nonlinear effects, such as exponential growth of sideband harmonics and formation of extreme amplitude wave structures -- rogue waves \cite{Zakharov2009, pelinovsky2008book, onorato2013rogue, dudley2019rogue}. 
Furthermore, adding a modulationally unstable background field to the system produces a new class of coherent structures called breathers, with an attractive rich physics and deep mathematical description \cite{akhmediev1997book, pelinovsky2008book}.
The model of the focusing one-dimensional nonlinear Schrödinger equation (fNLSE) in the presence of an unstable constant background has been successfully applied to describe real-world propagation and interactions of breathers in different physical waveguides
\cite{kibler2010, chabchoub2011, bailung2011, frisquet2013collision, kibler2015}. 
In our work, we write the fNLSE in the dimensionless form
\begin{eqnarray}\label{eqNLSdimentionless}
	i \psi_{t} + \frac12 \psi_{xx} + |\psi|^2 \psi = 0,
\end{eqnarray}
where $\psi(t,x)$ describes a complex-valued wave field. 
The evolution variable $t$ is time, while the spatial variable $x$ is coordinate. 
The role of the constant amplitude background plays the exact solution of  Eq.~(\ref{eqNLSdimentionless}),
\begin{equation}\label{condensate}
    \psi_0 = Ae^{i\Theta+iA^2 t},
\end{equation}
where $A$ and $\varphi$ are the real-valued background amplitude and phase. 
The family of fundamental breather solutions of the fNLSE includes famous exact solutions of Kuznetsov--Ma (KM) \cite{Kuznetsov1977, Ma1979}, Peregrine (P) \cite{peregrine1983}, Akhmediev (A) \cite{akhmediev1985generation}, and Tajiri-Watanabe (TW) \cite{tajiri1998breather}, which properties have been previously studied in detail \cite{akhmediev1997book,pelinovsky2008book}. 
The KM, TW, and P breathers are spatially localized, and the A breather is spatially periodic and localized in time. 
Meanwhile, the P breather represents a wave structure doubly localized in space and time.
Thanks to the inverse scattering transform (IST) method \cite{novikov1984theory, ablowitz2011nonlinearbook}, the fNLSE can be analytically integrated employing the so-called auxiliary Zakharov--Shabat (ZS) system \cite{Zakharov1972}. 
The scattering problem formulated for the ZS system provides fundamental information about the wave field scattering data (IST spectrum), which changes trivially along with the fNLSE evolution and thus represents a nonlinear analog of conventional linear Fourier harmonics. 
The latter fact gives an alternative name to the IST approach -- the nonlinear Fourier transform \cite{osborne2010book, ablowitz2011nonlinearbook}. 
Provided the scattering data is available, one can predict the evolution of the system and identify the physical characteristics of solitary structures in the wavefield, such as amplitude, velocity, phase, and spatial position. 
The IST approach can be divided into two steps -- direct and inverse scattering transforms, which can be performed analytically or using approximation techniques only for a small number of wavefields, see e.g. \cite{satsuma1974b,tovbis2000eigenvalue,mullyadzhanov2021solitons}, thus, motivating the development of accurate numerical tools.
The IST approach has its alternative called the dressing method \cite{novikov1984theory, zakharov1978relativistically}, also known as the Darboux transformation \cite{akhmediev1991extremely, matveev1991darboux}, which is advantageous for constructing multi-soliton and multi-breather solutions.
With the growing interest in the IST approach for the analysis and synthesis of nonlinear wave fields \cite{osborne1995inverse, osborne2010book, chekhovskoy2019nonlinear, suret2020nonlinear, slunyaev2021persistence,teutsch2022contribution}, development of numerical methods for direct and inverse scattering problems attract more attention \cite{burtsev1998numerical,vaibhav2018higher,mullyadzhanov2019direct, medvedev2020exponential, Aref2019, prins2021accurate} with applications to the Cauchy problem \cite{trogdon2012numerical, trogdon2013numerical}, information transmission \cite{yousefi2014information, turitsyn2017nonlinear, Frumin2017}, fiber Bragg grating synthesis \cite{Frumin2015}, etc.
In this work we present a numerical direct scattering transform (DST) algorithm for spatially localized KM and TW breathers, representing a generalization of the Boffetta--Osborne approach originally developed for solitons \cite{BofOsb1992}.
We consider the scattering problem for the ZS system with boundary conditions corresponding to the non-vanishing background field and find analytical relations between scattering coefficients and transfer matrix elements in order to identify breathers instead of solitons.
To compute the transfer matrix numerically, we use the standard second-order Boffetta--Osborne scheme \cite{BofOsb1992} together with recently proposed high-order numerical schemes based on the Magnus expansion \cite{mullyadzhanov2019direct, mullyadzhanov2021magnus}.
We construct localized single- and multi-breather solutions of the fNLSE using the dressing method and verify the developed numerical approach to be able to accurately recover a complete set of the scattering data providing the information about the amplitude, velocity, phase, and position of breathers.
%

%-------------------------------------------------------------------------------------------------------------
%-------------------------------------------------------------------------------------------------------------

\section{Scattering problem for breathers}
\label{Sec:II}

The IST method for the fNLSE integration is based on the scattering problem for the following linear auxiliary Zakharov--Shabat (ZS) system~\cite{Zakharov1972}
\begin{eqnarray}
	\mathbf{\Phi}_{x} &=& \begin{pmatrix} -i \lambda & \psi \\ -\psi^* & i \lambda \end{pmatrix}\mathbf{\Phi},
	\label{ZSsystem1}\\
	\mathbf{\Phi}_t &=& \begin{pmatrix}\ -i\lambda^2 + \frac{i}{2} |\psi|^2 & \lambda \psi + \frac{i}{2} \psi_x \\ -\lambda \psi^* + \frac{i}{2} \psi^*_x & i\lambda^2 - \frac{i}{2} |\psi |^2 \end{pmatrix} \mathbf{\Phi},
	\label{ZSsystem2}
\end{eqnarray}
where $\mathbf{\Phi}(x, t, \lambda) = (\phi_1,\phi_2)^{T}$ is a two-component vector wave function. 
According to the IST theory, the fNLSE (\ref{eqNLSdimentionless}) is obtained as a compatibility condition $\mathbf{\Phi}_{xt} = \mathbf{\Phi}_{tx}$. 
In Eqs.~(\ref{ZSsystem1}) and (\ref{ZSsystem2}), the fNLSE wave field $\psi$ plays the role of a potential, while $\lambda = \xi + i \eta$ is a complex-valued spectral parameter and the star stands for the complex conjugate.
Within the scattering problem we fix $t=0$ without loss of generality and consider the fNLSE wave field with the constant amplitude boundary conditions
\begin{equation}\label{condensate_boundary_cond}
    \psi \rightarrow A e^{i\Theta_{\pm}}, \quad\text{at}\quad x \rightarrow \pm\infty.
\end{equation}
In other words, the wave field at infinity tends to the background (\ref{condensate}) with different phases in a general case.
The solution of Eqs. (\ref{ZSsystem1}), (\ref{condensate_boundary_cond}) can be expressed as the sum of two independent vectors
\begin{equation}
    \mathbf{\Phi} = 
    c_1 \begin{pmatrix}\ 1 \\ p\,e^{-i\Theta_{\pm}} \end{pmatrix}e^{-i\zeta x} +
    c_2 \begin{pmatrix}\ p\, e^{i\Theta_{\pm}} \\ 1 \end{pmatrix}e^{i\zeta x},
\end{equation}
where $c_1$ and $c_2$ are arbitrary complex-valued constants and the functions $\zeta(\lambda)$ and $p(\lambda)$ are
\begin{eqnarray}\label{zeta}
    \zeta &=& \sqrt{\lambda^2 + A^2},
\\\label{rho}
    p &=& -\frac{iA}{\lambda + \zeta}.
\end{eqnarray}
The function $\zeta(\lambda)$ introduces two Riemann sheets of the spectral parameter plane with a branch cut, which we define on the line $[-iA,iA]$.
We choose the Riemann sheet with the following condition
\begin{eqnarray}\label{zeta_sign}
    &&\mathrm{Im}[\zeta]\geq 0, \quad\text{at}\quad \mathrm{Im}[\lambda] = \eta \geq 0,
    \\\nonumber
    &&\mathrm{Im}[\zeta]<0, \quad\text{at}\quad \mathrm{Im}[\lambda] = \eta <0.
\end{eqnarray}
By analogy with the formulations in quantum mechanics \cite{landau1958quantum} and derivations of the IST scheme \cite{Ma1979}, we introduce scattering coefficients $a(\lambda)$ and $b(\lambda)$ fixing the asymptotics of the wave function as
\begin{eqnarray}
\label{wave_function}
	\lim_{x\to -\infty}\biggl\{ \mathbf{\Phi} &-& \begin{pmatrix}\ 1 \\ p e^{-i\Theta_{-}} \end{pmatrix}e^{-i\zeta x}\biggr\} = 0, \label{ScatteringProblemAsymptotics}\\\nonumber
	\lim_{x \to +\infty}\biggl\{\mathbf{\Phi} &-& a\begin{pmatrix}\ 1 \\ p e^{-i\Theta_{+}} \end{pmatrix}e^{-i\zeta x} + 
    b \begin{pmatrix}\ p e^{i\Theta_{+}} \\ 1 \end{pmatrix}e^{i\zeta x}\biggr\} = 0.
\end{eqnarray}
Note that in contrast to \cite{Ma1979}, we consider a general case with different phases $\Theta_{\pm}$ of the background at $x \to \pm \infty$.
According to the IST theory \cite{Kuznetsov1977, Ma1979}, it is sufficient to consider only the upper half of the spectral parameter plane and only the sheet of the Riemann surface of the function $\zeta(\lambda)$ where relations (\ref{zeta_sign}) are fulfilled.
The eigenvalue spectrum of the scattering problem (\ref{ZSsystem1}) consists of the eigenvalues $\lambda$ corresponding to bounded solutions $\mathbf{\Phi}$ of the ZS systems with asymptotics~(\ref{ScatteringProblemAsymptotics}). 
Such solutions exist for the continuous sets of real-valued function $\zeta(\lambda)$, i.e., when $\lambda = \xi\in\mathbb{R}$ and $\lambda = i\eta,\, \eta\in(0,1]$. 
In addition, the bounded solutions exist for discrete complex-valued points $\lambda_j$ lying outside the branch cut if $a(\lambda_j) = 0$.
The latter part of the eigenvalue spectrum usually consists of a finite number of discrete points $j=1,...,N$ and is called a discrete spectrum. 
The eigenvalues lying on the real line and on the branch cut line form the stable and unstable parts of the continuous spectrum.

The full set of the scattering data represents a combination of the discrete $\{\lambda_j, \rho_j \}$, stable $\{ r_{\mathrm{s}} \}$ and unstable $\{ r_{\mathrm{MI}} \}$ continuous spectra; see \cite{Kuznetsov1977,Ma1979} for details,
\begin{eqnarray}
	&& \big\{
	\lambda_j \,\,|\,\, a(\lambda_j) = 0, \,\, \mathrm{Im}\,\lambda_j > 0
	\big\},\quad \bigg\{\rho_j = \frac{b(\lambda_j)}{a'(\lambda_j)}\bigg\},
    \nonumber\\
	&&
	r_{\mathrm{s}}(\xi) = \frac{b(\xi)}{a(\xi)}, \qquad
    r_{\mathrm{MI}}(\eta) = \frac{b(i\eta)}{a(i\eta)}, \qquad
    \xi\in\mathbb{R}, \qquad \eta\in(0,1]. \label{ScatteringData}
\end{eqnarray}
In (\ref{ScatteringData}) $a'(\lambda)$ represents complex derivative of $a(\lambda)$ with respect to $\lambda$, while $\rho_j$ are the so-called norming constants associated with the eigenvalues $\lambda_j$. 
One can find the wave field $\psi$ of the fNLSE by a known set of scattering data (\ref{ScatteringData}) solving the system of integral Gelfand--Levitan--Marchenko (GLM) equations. 
In the general case, this can only be done numerically, asymptotically at a large time or in the semi-classical approximation \cite{lewis1985semiclassical, jenkins2014semiclassical}.
Each discrete eigenvalue $\lambda_j$ corresponds to a breather in the wave field. 
We study only spatially localized KM and TW breathers, which are characterized by the following choices of the discrete spectral parameter:
\begin{eqnarray}
\label{KM_par_set}
&&\text{(KM)} \qquad \{\mathrm{Re}[\lambda] = 0, \,\, \mathrm{Im}[\lambda] < A\},
\\
\label{TW_par_set}
&&\text{(TW)} \qquad \{\mathrm{Re}[\lambda] \ne 0 \}.
\end{eqnarray}
In the pure discrete spectrum case, i.e., when the reflection coefficients are zero, i.e. $\{ r^{\mathrm{s}} \} = 0$, $\{ r^{\mathrm{MI}} \} = 0$, the IST procedure can be performed analytically solving the GLM equations, leading to an exact $N$-breather solution $\psi_{(N)}(x,t)$.
Also in this case the scattering coefficient $a(\lambda)$ can be found in the following explicit form, see \cite{Kuznetsov1977,Ma1979},
\begin{equation}\label{a_Nbreather}
    a_{(N)}(\lambda) = \prod_{j=1}^{N}
    \frac{\lambda - \lambda_j + \zeta - \zeta_j}{\lambda - \lambda^*_j + \zeta - \zeta^*_j}
    \times
    \frac{\lambda - \lambda^*_j + \zeta + \zeta^*_j}{\lambda - \lambda_j + \zeta + \zeta_j}.
\end{equation}

\section{Dressing method for breathers}
\label{Sec:III}

The conventional IST approach requires solving the GLM equations to construct exact multi-breather solutions of the fNLSE. Alternatively one can use a simplified version of the IST approach, called the dressing method (DM), also known as the Darboux transformation. 
The DM represents a recursive algebraic scheme~\cite{novikov1984theory,matveev1991darboux} for constructing exact solutions to integrable nonlinear partial differential equations, and is often used to find general multi-soliton and multi-breather solutions of the fNLSE~(\ref{eqNLSdimentionless}), see e.g.~\cite{zakharov1978relativistically, akhmediev1991extremely, akhmediev2009extreme, gelash2014superregular}.
The dressing procedure starts with choosing an exact solution to the fNLSE for which the exact solution to the ZS system is known as well. 
In case of breathers, we take the unperturbed background solution (\ref{condensate}). 
The corresponding matrix solution of the ZS system (\ref{ZSsystem1}) and (\ref{ZSsystem2}) is as follows,
\begin{eqnarray}\label{Psi0cond}
	&& \mathbf{\widehat{\Phi}}_{0}(x,t,\lambda) = \begin{pmatrix}\ e^{\phi_0-\phi}  &  p e^{\phi_0+\phi}  \\  p e^{-\phi_0-\phi}  & e^{-\phi_0+\phi} \end{pmatrix},
	\label{Psi0cond}\\
	&& \phi_0 = i\frac{A^2t +\Theta}{2}, \quad \phi = i\zeta(x + \lambda t).
	\nonumber
\end{eqnarray}

\begin{figure*}[t!]\centering
	\includegraphics[width=0.32\linewidth]{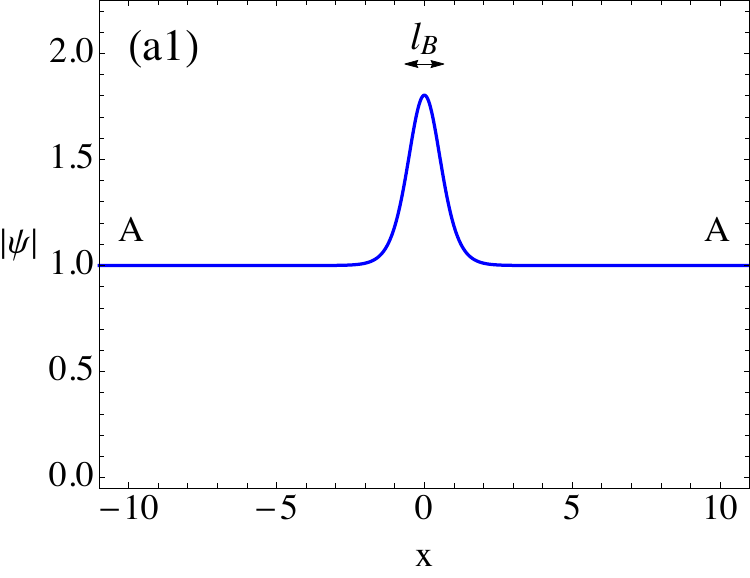}
	\includegraphics[width=0.32\linewidth]{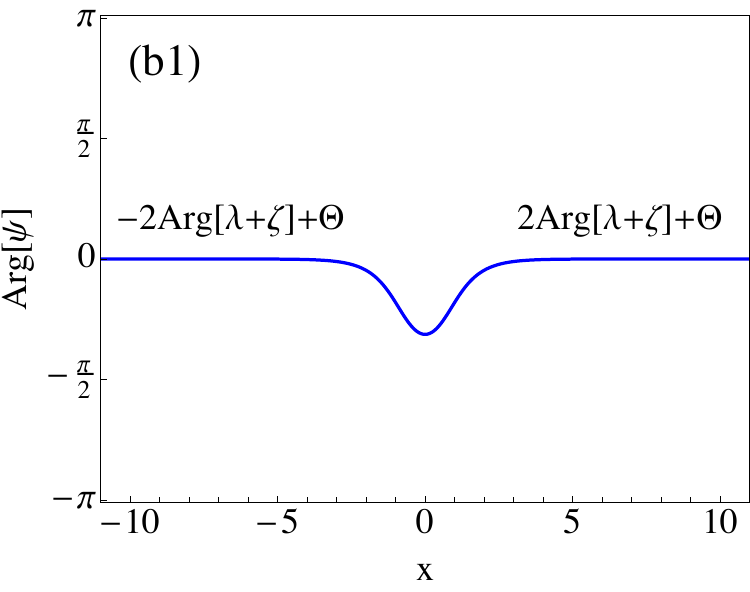}
    \includegraphics[width=0.32\linewidth]{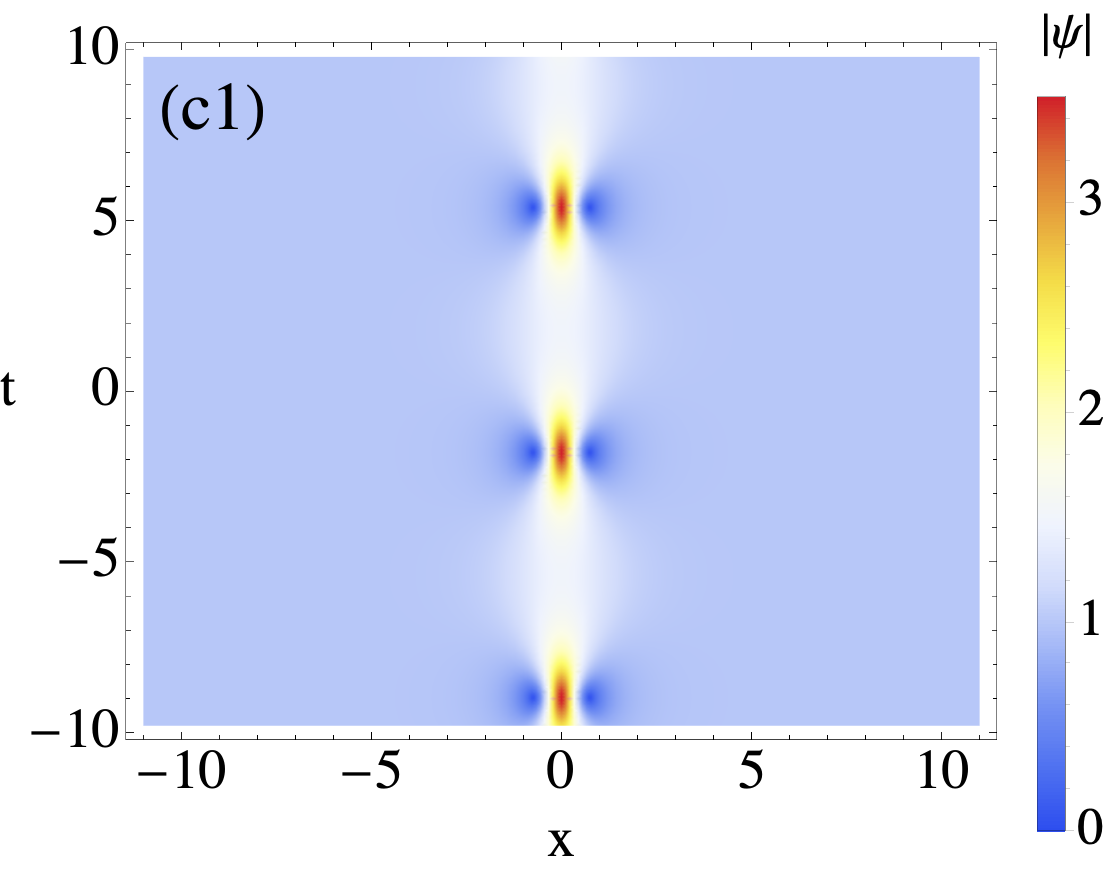}
\\
	\includegraphics[width=0.32\linewidth]{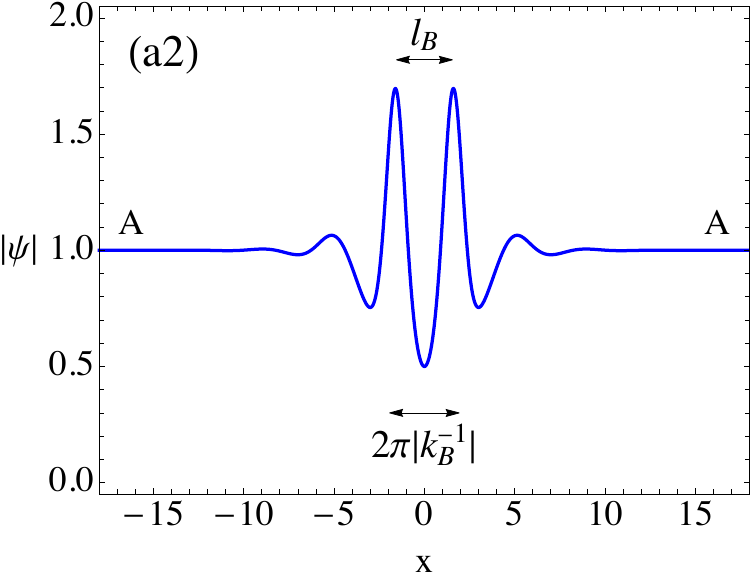}
	\includegraphics[width=0.32\linewidth]{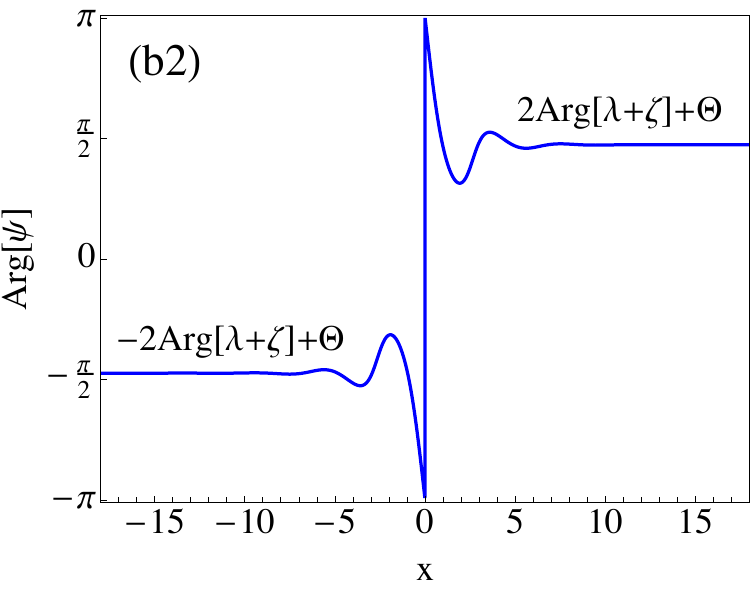}
    \includegraphics[width=0.32\linewidth]{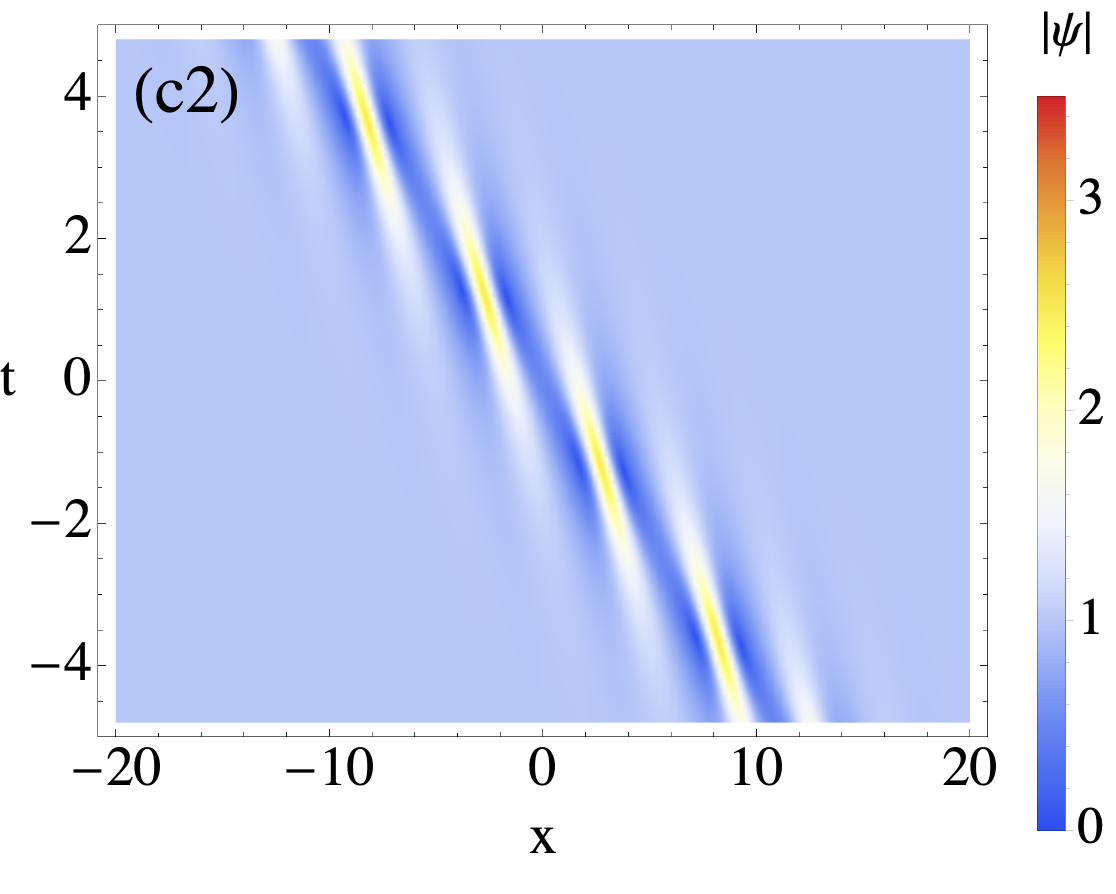}
	\caption{Behaviour and characteristics of spatially localized KM (top row) and TW (bottom row) breathers. (a) Wave field amplitude $|\psi(x)|$ at $t=0$ with indicated breather parameters (\ref{characteristic_values}) and wave field asymptotic values (\ref{1Bsolution_asympt_lambda}). (b) Wave field phase $\mathrm{Arg}[\psi(x)]$ at $t=0$ with indicated asymptotic values (\ref{1Bsolution_asympt_lambda}). (c) Spatio-temporal evolution of $|\psi|$. Here and in the next figures, the function $\mathrm{Arg}$ is defined in the cyclic interval $[-\pi,\pi)$, so that the function value can exhibit a jump as in panel (b2).
	}
	\label{fig:1}
\end{figure*}

At the $n$-th step of the recursive method, the $n$-breather potential $\psi_{(n)}(x)$ is constructed via the $(n-1)$-breather potential $\psi_{(n-1)}(x)$ and the corresponding matrix solution $\mathbf{\widehat{\Phi}}^{(n-1)}(x,\lambda)$ as
\begin{eqnarray}\label{psi_n}
	\psi_{(n)}(x) = \psi_{(n-1)}(x) + 2i(\lambda_n-\lambda^*_n)\frac{q^*_{n1}q_{n2}}{|\mathbf{q_n}|^2},
\end{eqnarray}
where the vector $\mathbf{q}_{n}=(q_{n1},q_{n2})^{\mathrm{T}}$ is determined by $\mathbf{\Phi}^{(n-1)}(x,\lambda)$ and the scattering data of the $n$-th breather $\{\lambda_{n},C_{n}\}$, 
\begin{eqnarray}\label{qn}
	\mathbf{q}_{n}(x) = [\mathbf{\widehat{\Phi}}^{(n-1)}(x,\lambda^*_n)]^{*}\times 
	\left(\begin{array}{c} 1 \\C_n \end{array}\right).
\end{eqnarray}
Here $C_n$, $n=1,...,N$, are the breather norming constants in the DM formalism, which describe breather phases and spatiotemporal locations; see details below.
The corresponding matrix solution $\mathbf{\widehat{\Phi}}^{(n)}(x,\lambda)$ of the ZS system is calculated via the so-called dressing matrix $\boldsymbol{\widehat{\sigma}}^{(n)}(x,\lambda)$
\begin{eqnarray}
	\mathbf{\widehat{\Phi}}^{(n)}(x,\lambda) &=& \boldsymbol{\widehat{\sigma}}^{(n)}(x,\lambda)\times \mathbf{\widehat{\Phi}}^{(n-1)}(x,\lambda), \label{dressing Psi}\\
	\boldsymbol{\widehat{\sigma}}^{(n)}_{ml}(x,\lambda) &=& \delta_{ml} + \frac{\lambda_n-\lambda_n^*}{\lambda - \lambda_n}\frac{q_{nm}^{*}q_{nl}}{|\mathbf{q_n}|^2}, \label{dressing matrix}
\end{eqnarray}
where $m,l=1,2$ and $\delta_{ml}$ is the Kronecker delta. 
Similar to the case of the fNLSE solitons on zero background, in the case of breathers the DM norming constants $C_n$ of the $N$-breather solution are related to the IST norming constants $\rho_n$ as follows; see e.g., \cite{aref2016control,gelash2020anomalous},
\begin{eqnarray}\label{rhok_Ck_conn}
	\rho_n = \frac{1}{C_n a'_{(N)}(\lambda_n)},  
\end{eqnarray}
where the derivative $a'_{(N)}(\lambda_n)$ can be evaluated analytically from Eq.~(\ref{a_Nbreather}) as
\begin{eqnarray}\label{aderiv_Nbreather}
    a'_{(N)}(\lambda_k) = 
    \frac{\lambda_k - \lambda^*_k + \zeta_k + \zeta^*_k}{\lambda_k - \lambda^*_k + \zeta_k - \zeta^*_k}
    \times
    \frac{1+\lambda_k/\zeta_k}{2\zeta_k}\times
    \prod_{j=1,l\ne k}^{N}
    \frac{\lambda_k - \lambda_j + \zeta_k - \zeta_j}{\lambda_k - \lambda^*_j + \zeta_k - \zeta^*_j}
    \cdot
    \frac{\lambda_k - \lambda^*_j + \zeta_k + \zeta^*_j}{\lambda_k - \lambda_j + \zeta_k + \zeta_j}.
\end{eqnarray}
We provide rigorous proof of the relation (\ref{rhok_Ck_conn}) in Appendix Sec.~\ref{Sec:Appendix1}.
In particular, one obtains as a result of the dressing procedure the following single-breather solution,
\begin{eqnarray}
	\psi_{(1)}(x,t) =  e^{iA^{2}t+i\Theta} \bigg(A - 4\eta_1 \frac{\tilde{q}_{1}^*\tilde{q}_{2}}{|\tilde{q}_{1}|^2+|\tilde{q}_{2}|^2} \bigg).
	\label{1Bsolution}
\end{eqnarray}
where the vector $\mathbf{\tilde{q}} = (\tilde{q}_{1}, \tilde{q}_{2})^{\mathrm{T}}$ has the components, 
\begin{eqnarray}
\tilde{q}_{1} = e^{\phi} - pCe^{-\phi},
\qquad
\tilde{q}_{2} = - p e^{\phi} + C e^{-\phi}.
\end{eqnarray}
Here we introduced a new vector with a tilde to extract a general phase multiplier from the solution (\ref{1Bsolution}).
The solution (\ref{1Bsolution}) with KM and TW eigenvalues (\ref{KM_par_set}), (\ref{TW_par_set}) describes a localized breather propagating on the top of the constant background with a characteristic size $l_{\mathrm{B}}$, propagation group velocity $V_{\mathrm{B}}$ and characteristic wave number of its internal spatial oscillations $k_{\mathrm{B}}$,
\begin{equation}
\label{characteristic_values}
    l_{\mathrm{B}} = (2 \mathrm{Im}[\zeta])^{-1}, 
    \qquad
    V_{\mathrm{B}} = \frac{\mathrm{Im}[\lambda\zeta]}{\mathrm{Im}[\zeta]},
    \qquad
    k_{\mathrm{B}} = 2 \mathrm{Re}[\zeta].
\end{equation}
Meanwhile, parametrizing the norming constant in the following way,
\begin{eqnarray}
	C = -\exp\big[2i\zeta x_{0} + i\theta\big]. \label{C_param_breathers}
\end{eqnarray}
we set the initial position of the breather $x_{0}$ and its phase $\theta_{0}$. Note that the parameters $x_{0}$ and $\theta_{0}$ are equal to the observed in the physical space position
and phase of a breather only for the one-breather solution. In the presence of other breathers or dispersive waves, the observed position and phase of a breather may differ considerably from the IST parameters $x_{0}$ and $\theta_{0}$.
The propagation velocity of the KM breather is zero, see (\ref{characteristic_values}), so it can be viewed as a standing case of the general TW breather. 
Fig.~\ref{fig:1} shows the wave field behavior and characteristics for the KM and TW breathers. 
More details on the breather dynamics theory can be found for example in monographs \cite{akhmediev1997book, pelinovsky2008book}.
It is important to note that for spatially localized breathers, the asymptotic background phase before and after the breather can be found as,
\begin{eqnarray}
\label{1Bsolution_asympt_lambda}
\psi_{(1)}  \ra A e^{\pm 2i\mathrm{Arg} [\zeta + \lambda ]+iA^{2}t+i\Theta} 
\quad\text{at}\quad x\rightarrow\pm\infty.
\end{eqnarray}
In particular, from (\ref{1Bsolution_asympt_lambda}) one sees that the TW breather has different condensate phases before and after itself; see Fig.~\ref{fig:1}(b2). 
In a general case of $N$ localized breathers, the background phase can be found as a multiplication of the factors (\ref{1Bsolution_asympt_lambda}) corresponding to each of the breathers; i.e.,
\begin{eqnarray}
\label{NBsolution_asympt_lambda}
\psi_{(N)} \ra A e^{\pm 2i\sum_{j=1}^{j=N} \mathrm{Arg}[(\zeta_j + \lambda_j) ]+iA^{2}t+i\Theta} 
\quad \text{at} \quad x\rightarrow\pm\infty,
\end{eqnarray}
Note that adding each KM-breather with $\lambda = i\eta$, $\eta>A$, see Eq.~(\ref{KM_par_set}) to the wave field, changes the solution sign since $2\mathrm{Arg}[\zeta(i\eta)+i\eta] = \pi$, see Eq.~(\ref{1Bsolution_asympt_lambda}) and Eq.~(\ref{zeta_sign}). 
The latter is a mathematical feature of the DM and the general solution phase can always be adjusted as needed using the factor $\Theta$.

\section{Direct scattering transform algorithm}
\label{Sec:IV}

In this section, we introduce an algorithmic realization of the DST in the presence of the constant background (\ref{condensate}) in order to identify numerically the full set of scattering data (\ref{ScatteringData}).
The novel approach represents a generalization of the Boffetta--Osborne algorithm for zero background \cite{BofOsb1992}.
As a first step we consider the problem on a finite computational domain $x \in \bigl[-L, L \bigr]$ instead of an infinite one shifting the boundary conditions (\ref{wave_function}) from $x \to \pm \infty$ to $x = \pm L$.
The potential $\psi(x)$ has to be well localized within $x \in \bigl[-L, L \bigr]$ tending to asymptotical values (\ref{condensate_boundary_cond}) at the boundaries.
We introduce a truncated wave function $\Phi_{\mathrm{tr}}$ and scattering coefficients $a_{\mathrm{tr}}(\lambda)$, $b_{\mathrm{tr}}(\lambda)$, corresponding to the shifted boundary conditions:
\begin{equation}
\label{ScatteringProblemTrunct}
\bold{\Phi}_{\mathrm{tr}}(-L) = 
\begin{pmatrix}\ 1 \\ p e^{-i\Theta_{-}} \end{pmatrix}e^{i\zeta L}
, \qquad
\bold{\Phi}_{\mathrm{tr}}(L) = 
a_{\mathrm{tr}}\begin{pmatrix}\ 1 \\ p e^{-i\Theta_{+}} \end{pmatrix}e^{-i\zeta L} + 
b_{\mathrm{tr}}\begin{pmatrix}\ p e^{i\Theta_{+}} \\ 1 \end{pmatrix}e^{i\zeta L}.
\end{equation}
For a finite computational domain the scattering data is expressed as follows:
\begin{eqnarray}
	&& \big\{
	\lambda^{\mathrm{tr}}_j \,\,|\,\, a_{\mathrm{tr}}(\lambda^{\mathrm{tr}}_j) = 0, \quad \mathrm{Im}\lambda^{\mathrm{tr}}_j > 0
	\big\},\quad \bigg\{\rho^{\mathrm{tr}}_{j} = \frac{b_{\mathrm{tr}}(\lambda^{\mathrm{tr}}_j)}{a'_{\mathrm{tr}}(\lambda^{\mathrm{tr}}_j)}\bigg\},
    \nonumber\\
	&&
	r^{\mathrm{s}}_{\mathrm{tr}}(\xi) = \frac{b_{\mathrm{tr}}(\xi)}{a_{\mathrm{tr}}(\xi)}, \qquad
    r_{\mathrm{tr}}^{\mathrm{MI}}(\eta) = \frac{b_{\mathrm{tr}}(i\eta)}{a_{\mathrm{tr}}(i\eta)}, \qquad
    \xi\in\mathbb{R}, \qquad \eta\in(0,1].
    \label{ScatteringData_truncated}
\end{eqnarray}
%
%which asymptotically reach the exact ones (\ref{ScatteringData}) for $L \to \infty$.
%

%
%
Then we construct the four component function $(\bold{\Phi}_{\mathrm{tr}},\bold{\Phi'}_{\mathrm{tr}})^T$ and introduce the $4\times 4$ transfer matrix $\bold{\widehat{T}}$ so that,
\begin{equation}
\label{T_matrix}
\begin{pmatrix}
    \bold{\Phi}_{\mathrm{tr}} (L,\lambda) \\
    \bold{\Phi'}_{\mathrm{tr}} (L,\lambda)
\end{pmatrix}
=
\underbrace{
\begin{pmatrix}
\bold{\widehat{\Sigma}} & 0\\
 \bold{\widehat{\Sigma}}' & \bold{\widehat{\Sigma}}
\end{pmatrix}
}_{\bold{\widehat{T}}}
\begin{pmatrix}
    \bold{\Phi}_{\mathrm{tr}} (-L,\lambda) \\
    \bold{\Phi'}_{\mathrm{tr}} (-L,\lambda)
\end{pmatrix}.
\end{equation}
Here $\bold{\widehat{\Sigma}}(\lambda)$ is a $2 \times 2$ transfer matrix for $\bold{\Phi}_{\mathrm{tr}}$, i.e. $\bold{\Phi}_{\mathrm{tr}}(L) = \bold{\widehat{\Sigma}} \bold{\Phi}_{\mathrm{tr}}(-L)$. 
In other words the role of $\bold{\widehat{T}}$ and $\bold{\widehat{\Sigma}}$ is to transfer the corresponding wave functions from $-L$ to $L$. 
The derivatives in (\ref{T_matrix}) are with respect to $\lambda$, i.e., $\bold{\Phi'}_{\mathrm{tr}} = \partial_{\lambda} \bold{\Phi}_{\mathrm{tr}}$ and $ \bold{\widehat{\Sigma}}'=  \partial_{\lambda}\bold{\widehat{\Sigma}}$.
The transfer matrix $\bold{\widehat{T}}(\lambda)$ contains the information about the scattering coefficients $a_{\mathrm{tr}}(\lambda)$, $b_{\mathrm{tr}}(\lambda)$ and their derivatives. 
We consider the asymptotic conditions (\ref{ScatteringProblemTrunct}) and using the definition (\ref{T_matrix}) find the following relations between the truncated scattering coefficients and the elements $T_{m,l}$ of $\bold{\widehat{T}}$, where $m,l=1,2,3,4$,
\begin{eqnarray}
\label{a_b_via_S}
&&a_{\mathrm{tr}}(\lambda) = \frac{e^{2 i \zeta L}}{1-p^2} 
\biggl\{ T_{11} + T_{12}pe^{-i\Theta_{-}}
- T_{21} p e^{i\Theta_{+}} - T_{22}p^2 e^{i(\Theta_{+} - \Theta_{-})} \biggr\},
\\\nonumber
&&b_{\mathrm{tr}}(\lambda) = \frac{1}{1-p^2} 
\biggl\{ -T_{11}pe^{-i\Theta_{+}} - T_{12}p^2e^{-i(\Theta_{-}+\Theta_{+})}
+ T_{21} + T_{22}p e^{-i\Theta_{-}} \biggr\}.
\end{eqnarray}
In a similar way, we find the derivative $a_{\mathrm{tr}}(\lambda)$ as,
\begin{eqnarray}
\label{da_S}
&&a'_{\mathrm{tr}}(\lambda) = \frac{e^{2 i \zeta L}}{1-p^2} 
\biggl( T_{31} + T_{32}pe^{-i\Theta_{-}} - T_{41}pe^{i\Theta_{+}}
- T_{42}p^2e^{i(\Theta_{+}-\Theta_{-})} 
\\\nonumber
&& 
+ i\frac{L\lambda}{\zeta} \biggl\{ T_{11} + T_{12}pe^{-i\Theta_{-}}
- T_{21}pe^{i\Theta_{+}} - T_{22}p^2 e^{i(\Theta_{+} - \Theta_{-})} + T_{33}
- T_{43}pe^{i\Theta_{+}} +T_{34}pe^{-i\Theta_{-}} - T_{44}p^2 e^{i(\Theta_{+} - \Theta_{-})}
\biggr\}
\\\nonumber
&& + \frac{p}{\zeta (1-p^2)} \biggl\{ -2T_{11}p - 2 T_{12}p^2 e^{-i\Theta_{-}} 
+ T_{21}(1+p^2)e^{i\Theta_{+}} 
\\\nonumber
&& +T_{22}p(1+p^2)e^{i(\Theta_{+}-\Theta_{-})}
-T_{34}(1-p^2)e^{-i\Theta_{-}} +T_{44}pe^{i(\Theta_{+}-\Theta_{-})} \biggr\}\biggr).
\end{eqnarray}
The numerical realization of the DST boils down to the discretization of the interval $[-L,L]$ into $M$ bins and computation of the $\bold{\widehat{T}}$ using a finite-difference scheme.
At the final stage, one retrieves numerically the scattering data set $\{ \lambda^{\mathrm{num}}_n,\rho^{\mathrm{num}}_n, r^{\mathrm{s}}_{\mathrm{num}},r^{\mathrm{MI}}_{\mathrm{num}} \}$ using the relations (\ref{a_b_via_S}) and (\ref{da_S}).
To compute $\bold{\widehat{T}}$ we use the Boffetta--Osborne second-order scheme \cite{BofOsb1992} together with recently proposed fourth- and sixth-order schemes based on the Magnus expansion \cite{mullyadzhanov2019direct, mullyadzhanov2021magnus}.
We reproduce these numerical schemes in full in Appendix Sec. \ref{Sec:Appendix2}.
Note that here we focus only on the discrete part of the scattering data corresponding to breathers, and leave the continuous spectrum part for further studies.

\section{Numerical examples}
\label{Sec:V}

\begin{figure*}[t!]\centering
	\includegraphics[width=0.32\linewidth]{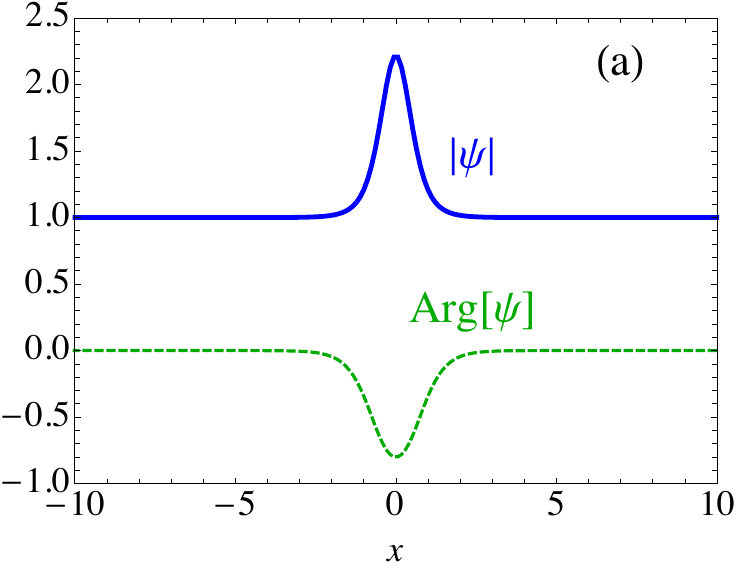}
	\includegraphics[width=0.32\linewidth]{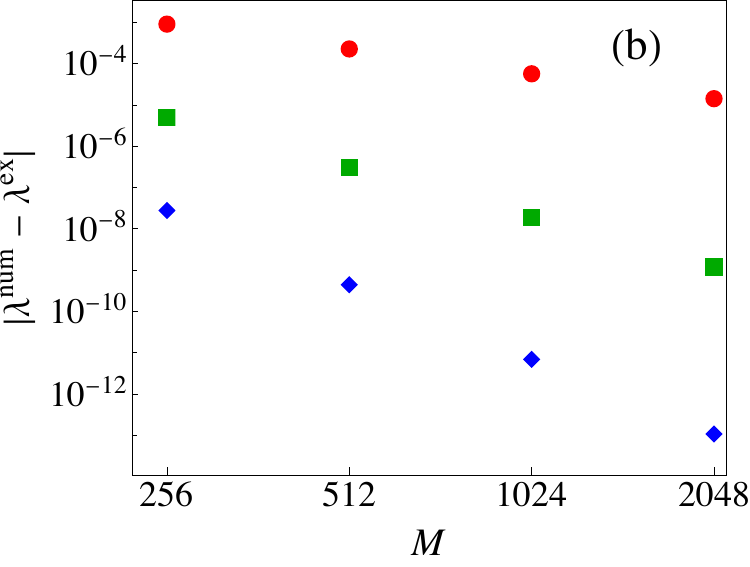}
 \includegraphics[width=0.32\linewidth]{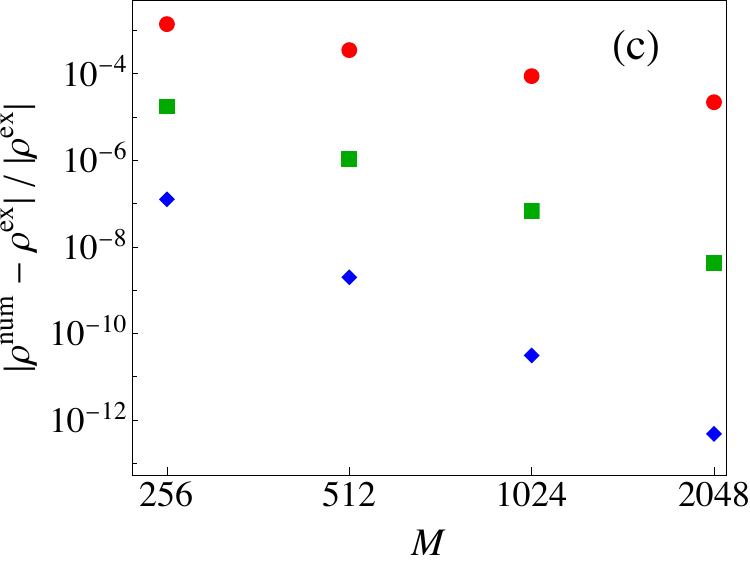}
	\caption{\small Convergence test of the developed DST algorithm. 
    The wave field is represented by a single KM breather shown in (a). 
    Panels (b) and (c) demonstrate the absolute value of the eigenvalue identification and the relative error for norming constant identification. 
    Red dots, green squares, and blue diamonds in (b) and (c) correspond to the second-, fourth-, and six-order of the scheme convergence.
	}
	\label{fig:2}
\end{figure*}

\begin{figure*}[t!]\centering
	\includegraphics[width=0.32\linewidth]{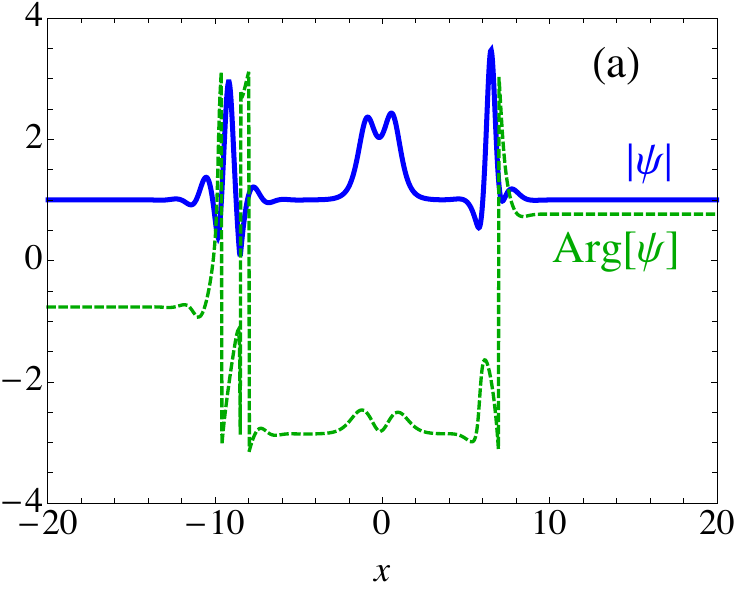}
	\includegraphics[width=0.32\linewidth]{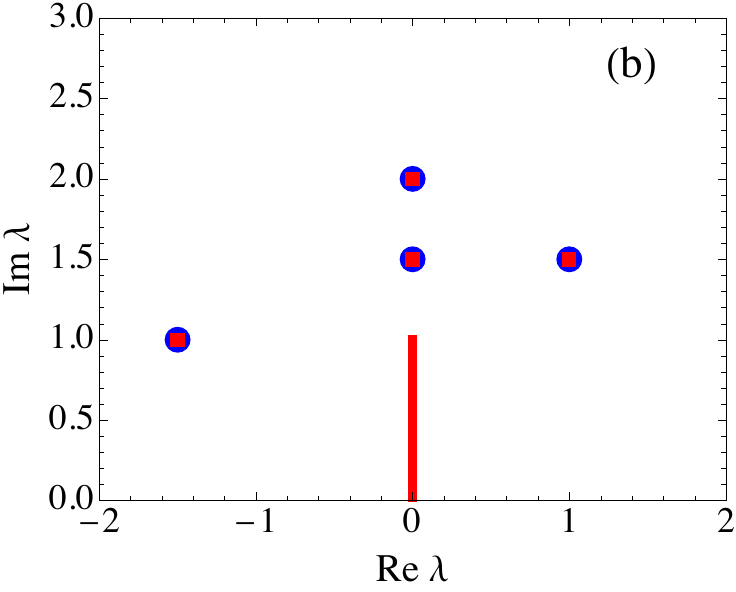}
 \includegraphics[width=0.32\linewidth]{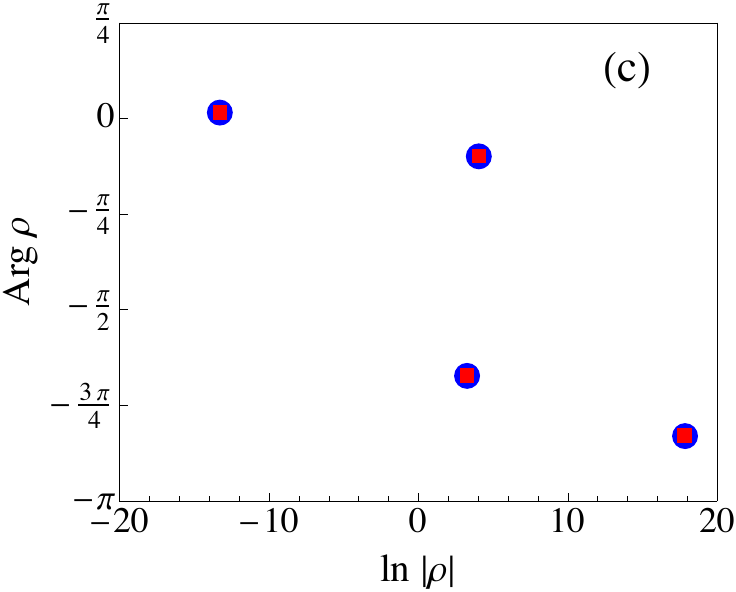}
	\caption{\small 
    Verification of the developed DST algorithm using a multi-breather complex example. 
    The wave field shown in panel (a) is represented by two KM breathers located at the center and two TW breathers on the left and right sides of the computational domain. 
    Panel (b) demonstrate the $\lambda$-plane of spectral parameter with the branchcut (red solid line), $\lambda^{\mathrm{ex}}$ (blue circles) and $\lambda^{\mathrm{num}}$ (red squares). 
    Panel (c) shows $\rho^{\mathrm{ex}}$ (blue circles) and $\rho^{\mathrm{num}}$ (red squares).
	}
	\label{fig:3}
\end{figure*}

To verify the proposed algorithm, we consider one example of a single KM breather and another example of a four-breather complex of KM and TW breathers. 
We use the dressing method to generate wave fields, and then the direct scattering transform to identify the corresponding scattering data, employing relations (\ref{rhok_Ck_conn}) connecting the DM and DST approaches. 
More precisely, (i) we set the initial exact scattering data of the breathers $\{ \lambda^{\mathrm{ex}}_n,\rho^{\mathrm{ex}}_n \}$, (ii) transform $\rho^{\mathrm{ex}}_n$ to $C^{\mathrm{ex}}_n$ using relation (\ref{rhok_Ck_conn}), and (iii) obtain the corresponding breather wave fields with the help of the dressing method scheme. 
Finally, we run the DST algorithm described in the previous section, see also Appendix \ref{Sec:Appendix2}, to identify the scattering data set $\{ \lambda^{\mathrm{num}}_n,\rho^{\mathrm{num}}_n \}$ numerically.
Fig.~\ref{fig:2} shows the convergence tests of our second-, fourth-, and sixth-order DST schemes applied to a single KM breather wave field with $\lambda^{\mathrm{ex}}=1.5 i$, $x^{\mathrm{ex}}_0=0$, $\theta^{\mathrm{ex}}=2.0$ and $\Theta=\pi$, see also Eq.~(\ref{C_param_breathers}). 
Fig.~\ref{fig:2}b,c demonstrate the absolute error of the eigenvalue identification $|\lambda^{\mathrm{num}}-\lambda^{\mathrm{ex}}|$ and the relative error of the norming constant identification $|\rho^{\mathrm{num}}-\rho^{\mathrm{ex}}|/|\rho^{\mathrm{ex}}|$, versus the number of the grid points $M$. 
The error data converges with slopes corresponding to the declared orders of the schemes used. 
Then Fig.~\ref{fig:3} shows the verification of the DST algorithm using a four-breather complex example composed of two KM and two TW breathers located at different areas of the computational domain. 
Parameters of breathers are set in such a way to model a general situation when the condensate phases $\Theta_{\pm}$ are different and as such nontrivially enter to the formulas (\ref{a_b_via_S}), (\ref{da_S}). 
The recovered scattering data of breathers $\{ \lambda^{\mathrm{ex}}_n,\rho^{\mathrm{ex}}_n \}$ are in precise correspondence with their exact counterpart $\{ \lambda^{\mathrm{num}}_n,\rho^{\mathrm{num}}_n \}$, that provides complete information about types of breathes, their characteristic parameters (\ref{characteristic_values}) and phase-position parameters (\ref{C_param_breathers}). 
Note that to recover the norming constants of the multi-breather complex accurately, we strength our algorithm with $100$-digits precision arithmetic. 
The latter is necessary to mitigate round-off and the so-called anomalous errors of the DST procedure; see our resent works elaborating on these issues \cite{gelash2020anomalous, agafontsev2023bound}.

\section{Conclusions}
\label{Sec:VI}

In the present paper, we generalize the classical Boffetta–Osborne DST algorithm \cite{BofOsb1992}, developed originally for solitons, to the field of breathers. The key ingredient of our approach represents the relations (\ref{a_b_via_S}) and (\ref{da_S}) between the scattering coefficients functions $a(\lambda)$, $b(\lambda)$, $a'(\lambda)$ and the elements of the $4\times 4$ transfer matrix $\bold{\widehat{T}}$ derived in the presence of the constant wave field background (\ref{condensate}). The derivation is based on the asymptotic behavior of the wave function (\ref{wave_function}) for spatially localized KM and TW breathers. To evaluate the scattering matrix numerically, we use the standard second-order approach from \cite{BofOsb1992} and the recently developed advanced high-order numerical schemes based on the Magnus expansion \cite{mullyadzhanov2019direct,mullyadzhanov2021magnus}. Our numerical tests demonstrate accurate recovery of the breather's parameters and confirm the declared convergence orders of the  developed schemes. Note that identifying the complete discrete scattering data set $\{ \lambda_n, \rho_n\}$ always requires consideration of the full scattering problem as we present here. Meanwhile computing solely the breathers eigenvalues $\{ \lambda_n \}$ can be performed using other techniques \cite{conforti2018auto} such as for example Fourier collocation method \cite{YangBook2010}. In our approach, we consider the general case of arbitrary asymptotic condensate phases $\Theta_{\pm}$ and thereby cover the whole general class of the localized breathers, except the P breather and its high-order analogs \cite{akhmediev2009rogue}. The first and high-order P breathers being a degenerate case of the KM class require a separate consideration.
%The latter is not necessary because in numerical DST the P breather is seen as KM breather with an eigenvalue closely located to the branch point due to the discrete approximation of the wave field, see also 

Our work contributes to the domain of numerical DST tools and suggests an efficient way to analyze the nonlinear dynamics of wave fields containing breathers. The latter is of special importance in light of the problem of modulation instability development, where breathers play one of the critical roles \cite{akhmediev1997book,zakharov2013nonlinear,kibler2015}. In particular, our DST algorithm can be used for a comprehensive analysis of breathers embedded into localized arbitrary-shaped background perturbations \cite{conforti2018auto}. With the complete set of scattering data, one can predict the picture of wave field evolution and also compute the pure breather's part of the wave field, elucidating their impact to the initial condition similar to as it was recently done with solitons \cite{gelash2021solitonic,agafontsev2023bound}. In addition one can use our approach to compute the reflection coefficients $\{ r_{\mathrm{s}},\,r_{\mathrm{MI}} \}$, which is important for problems where continuous spectrum waves dominate in the development of modulation instability or are in interplay with breathers \cite{biondini2016universal,conforti2018auto}. Another important direction represents the rapidly growing field of integrable turbulence, where the DST algorithms can be used to study phase correlations of individual breathers as well as phase correlations in gases of breathers \cite{zakharov2009turbulence,pelinovsky2013two,walczak2015optical, agafontsev2015integrable,soto2016integrable,suret2016single,roberti2021numerical,agafontsev2023bound}. Note that both IST and DST algorithms face rapid growth of numerical errors when increasing the number of discrete spectrum components, which can be coped with using high-precision arithmetic as recently suggested in \cite{gelash2018strongly,gelash2020anomalous,agafontsev2023bound}. We highlight that the complete characterization of the breather's parameters benefits the IST analysis of experimental data on coherent structures in optics, hydrodynamics, and other nonlinear waveguides described by nearly integrable models \cite{KivsharRMP1989,osborne2010book,suret2020nonlinear,randoux2018nonlinear,teutsch2022contribution}. Also our approach can be generalized to the case of vector breathers of the Manakov system \cite{baronio2014vector,kraus2015focusing,che2022nondegenerate,gelash2023vector}.

\section{Acknowledgements}

The work of IM on Section III was supported by the Russian Science Foundation (Grant No. 19-72-30028-$\Pi$).
The work of IM on Section V was supported by the Russian Science Foundation (Grant No. 22-22-00653).
The work of RM on Section IV and IX was supported by the Russian Science Foundation (Grant No. 19-79-30075-$\Pi$).
The work of AG was funded by the European Union's Horizon 2020 research and innovation program under the Marie Skłodowska-Curie grant agreement No. 101033047. 

\section{Appendix}

\subsection{Relation between IST and DM norming constants}
\label{Sec:Appendix1}

In this section, we prove the relation (\ref{rhok_Ck_conn}), which allows us to benefit from combining the IST and DM approaches in our work. First, we note that the matrix $\mathbf{\widehat{\Phi}}_{(n)}$ appearing at each step of the DM procedure represents in general two independent vector solutions of the ZS system with the potential $\psi_{n}$. These solutions are also called Jost functions \cite{novikov1984theory}. We denote them as $\mathbf{\Phi}_{(n),\mathrm{I}}$ and $\mathbf{\Phi}_{(n),\mathrm{II}}$ as shown in the following illustration,
\begin{eqnarray}
\label{MatrixPhi_as2vectors}
\mathbf{\widehat{\Phi}}_{(n)}=
\left (
\begin{array}{r|r}
\,\widehat{\Phi}_{(n),11} \, & \,\widehat{\Phi}_{(n),12} \, \\
\undermat{\mathbf{\Phi}_{(n),\mathrm{I}}}{\,\widehat{\Phi}_{(n),21}\,} & 
\undermat{\mathbf{\Phi}_{(n),\mathrm{II}}}{\,\widehat{\Phi}_{(n),22}\,} \\
\end{array}
\right ).
\\\nonumber
\end{eqnarray}
\newline

Any two solutions of the ZS system $(f_1,f_2)^{T}$ and $(g_1,g_2)^{T}$ are connected with each other via the so-called involution property; see e.g., \cite{novikov1984theory} for details, as,
\begin{equation}
\label{involution_general}
\begin{pmatrix}
    g_1(\lambda) \\
    g_2(\lambda)
\end{pmatrix}
=
\mathrm{Inv}\biggl [\begin{pmatrix}
    f_1(\lambda) \\
    f_2(\lambda)
\end{pmatrix} \biggr ]
=
\alpha\begin{pmatrix}
    -f_2^* (\lambda^*) \\
    f^*_1 (\lambda^*)
\end{pmatrix},
\end{equation}
where $\alpha$ is a constant. In particular the components of $\mathbf{\widehat{\Phi}}_{(n)}$ are in involution (\ref{involution_general}) with $\alpha=1$, i.e.,
\begin{equation}
\label{involution}
    \mathbf{\Phi}_{(n),\mathrm{II}} = \mathrm{Inv}\bigl [ \mathbf{\Phi}_{(n),\mathrm{I}} \bigr ].
\end{equation}
The value of $\alpha$ for (\ref{involution}) can be retrieved directly from the DM construction described in Sec.~\ref{Sec:III}. Indeed, for the matrix $\mathbf{\widehat{\Phi}}_{0}$ the value $\alpha=1$ follows from the Eq.~(\ref{Psi0cond}), while the subsequent steps of the dressing do not change it. 

According to the IST theory at the points of the discrete eigenvalues, $\lambda_n$ the two solutions of the ZS system become linearly dependent so that the ratio between them is constant at any spatial point \cite{novikov1984theory}. The value of the constant in general is arbitrary and depends on the solution's normalization. Considering similar to as it was done in \cite{gelash2020anomalous} the asymptotics of $\mathbf{\Phi}_{(n),\mathrm{I}}$ and $\mathbf{\Phi}_{(n),\mathrm{II}}$ at $x\ra-\infty$ and taken into account that $\alpha=1$ in the involution condition (\ref{involution}) we conclude that this ratio is the same as for the conventional Jost functions in the presence of the condensate background \cite{Ma1979}, namely,
\begin{equation}
\label{PsiI_and_PsiII_ratio}
    \mathbf{\Phi}_{(n),\mathrm{I}}(\lambda_n) = b(\lambda_n) \mathbf{\Phi}_{(n),\mathrm{II}}(\lambda_n).
\end{equation}

Since Eq.~(\ref{PsiI_and_PsiII_ratio}) is valid for any space point, we consider the simplest case $x=0$. Then at the first step of the dressing with $\{\lambda_1, C_1\}$ we introduce the value $p_1 = p(\lambda_1)$ and obtain,
\begin{equation}
\label{Psi_1_at_0}
    \mathbf{\widehat{\Phi}}_{(1)}(\lambda_1)|_{x=0} = \begin{pmatrix}\ (1-p^*_1 C^*_1)(1-p^2_1)  &  C_1 (1-p^*_1 C^*_1)(1-p^2_1)  \\  (C^*_1-p^*_1)(1-p^2_1)  & C_1 (C^*_1-p^*_1)(1-p^2_1) \end{pmatrix},
\end{equation}
meaning that $\mathbf{\Phi}_{(1),\mathrm{I}}(\lambda_n) = (1/C_1) \mathbf{\Phi}_{(1),\mathrm{II}}$. The subsequent steps of the dressing procedure at the point $\lambda=\lambda_1$ do not change this ratio; see Eq.~(\ref{dressing matrix}). The freedom to order the eigenvalues in the dressing method allows us to obtain relation (\ref{Psi_1_at_0}) for any $\lambda_n$, so that for any step $(n)$,
\begin{equation}
    \mathbf{\Phi}_{(n),\mathrm{I}}(\lambda_n) = \frac{\mathbf{\Phi}_{(n),\mathrm{II}}(\lambda_n)}{C_n} ,
\end{equation}
and thus,
\begin{equation}
    b_n = 1/C_n,
\end{equation}
that together with the definition of $\rho_n$; see Eq.~(\ref{ScatteringData}), proves the relation (\ref{rhok_Ck_conn}) between the IST and the DM norming constants.

\section{Numerical computation of transfer matrix}
\label{Sec:Appendix2}

Here we briefly describe the derivation of a class of high-order numerical schemes based on the Magnus expansion \cite{mullyadzhanov2019direct, mullyadzhanov2021magnus}.
Building finite-difference schemes starts with the discretization of the interval $[-L, L]$ into $M$ bins which may represent an unstructured mesh in a general case.
Let us denote the center of $m$th bin as $x_m$ with the width of $\Delta x_m$.
Using the Magnus expansion \cite{Magnus2009} the solution of the matrix first-order differential equation for Eq. (\ref{ZSsystem1}) within one bin can be written as follows:
\begin{eqnarray}
\label{MagnusRel} \mathbf{\Phi}(x_p) = \widehat{\mathbf{U}}(x_m) \mathbf{\Phi}(x_c), \;\;\;  \widehat{\mathbf{U}}(x_m) = e^{\widehat{\mathbf{\Omega}} (x_m)},
\end{eqnarray}
where in general $\widehat{\mathbf{\Omega}}$ represents an infinite series:
\begin{eqnarray}
\widehat{\mathbf{\Omega}} (x_m) = \sum_{j=1}^{\infty} \widehat{\mathbf{\Omega}}_j (x_m),
\label{MagnusOmega}
\end{eqnarray}
with
\begin{eqnarray}
\label{Om1} && \widehat{\mathbf{\Omega}}_1 (x_m) = \int_{x_c}^{x_p} \widehat{\mathbf{Q}}(x) dx, \\
\label{Om2} && \widehat{\mathbf{\Omega}}_2 (x_m) = \frac{1}{2} \int_{x_c}^{x_p} dx_1 \int_{x_c}^{x_1} dx_2 \{ \widehat{\mathbf{Q}}_1, \widehat{\mathbf{Q}}_2 \}, \\
\label{Om3} && \widehat{\mathbf{\Omega}}_3 (x_m) = \frac{1}{6} \int_{x_c}^{x_p} dx_1 \int_{x_c}^{x_1} dx_2 \int_{x_c}^{x_2} dx_3 \Big( \{ \widehat{\mathbf{Q}}_1, \{ \widehat{\mathbf{Q}}_2, \widehat{\mathbf{Q}}_3 \} \} + \{ \widehat{\mathbf{Q}}_3, \{ \widehat{\mathbf{Q}}_2, \widehat{\mathbf{Q}}_1 \} \} \Big), \\ 
%
%&& \times \Big( \{ Q_1, \{ Q_2, Q_3 \} \} + \{ Q_3, \{ Q_2, Q_1 \} \} \Big), \nonumber\\
%
\label{Om4} && \widehat{\mathbf{\Omega}}_4 (x_m) = \frac{1}{12} \int_{x_c}^{x_p} dx_1 \int_{x_c}^{x_1} dx_2 \int_{x_c}^{x_2} dx_3 \int_{x_c}^{x_3} dx_4  \Big( \{ \{ \{ \widehat{\mathbf{Q}}_1, \widehat{\mathbf{Q}}_2 \}, \widehat{\mathbf{Q}}_3 \}, \widehat{\mathbf{Q}}_4 \} + \{ \widehat{\mathbf{Q}}_1, \{ \{ \widehat{\mathbf{Q}}_2, \widehat{\mathbf{Q}}_3 \}, \widehat{\mathbf{Q}}_4 \} \} + \\
%
%&& \times \Big( \{ \{ \{ Q_1, Q_2 \}, Q_3 \}, Q_4 \} + \{ Q_1, \{ \{ Q_2, Q_3 \}, Q_4 \} \} + \nonumber\\
%
&& + \{ \widehat{\mathbf{Q}}_1, \{ \widehat{\mathbf{Q}}_2, \{ \widehat{\mathbf{Q}}_3, \widehat{\mathbf{Q}}_4 \} \} \} + \{ \widehat{\mathbf{Q}}_2, \{ \widehat{\mathbf{Q}}_3, \{ \widehat{\mathbf{Q}}_4, \widehat{\mathbf{Q}}_1 \} \} \} \Big), \nonumber
\end{eqnarray}
where $x_c = x_m - \Delta x_m / 2$, $x_p = x_m + \Delta x_m / 2$, $\{ \widehat{\mathbf{A}}, \widehat{\mathbf{B}}\} = \widehat{\mathbf{A}} \widehat{\mathbf{B}} - \widehat{\mathbf{B}} \widehat{\mathbf{A}}$ is the matrix commutator and $\widehat{\mathbf{Q}}(x_k) = \widehat{\mathbf{Q}}_k$ with the matrix appearing in Eq. (\ref{ZSsystem1}) as follows:
\begin{eqnarray}	
    \widehat{\mathbf{Q}} &=& \begin{pmatrix} -i \lambda & \psi \\ -\psi^* & i \lambda \end{pmatrix}.
	\label{Qmat}
\end{eqnarray}
The relation (\ref{MagnusRel}) allows to express the matrix $\bold{\widehat{\Sigma}}$:
\begin{eqnarray}
\label{SigmaMat} \bold{\widehat{\Sigma}} = \prod_{m=1}^{M} \widehat{\mathbf{U}} (x_m) = \prod_{m=1}^{M} e^{\widehat{\mathbf{\Omega}} (x_m)},
\end{eqnarray}
providing the way to compute matrix $\bold{\widehat{T}}$ according to Eq. (\ref{T_matrix}).
Another important ingredient is the polynomial representation of $\psi(x)$ inside the bin which can be performed using the Taylor series: % (see \textcolor{violet}{Supplement 1} for Lagrange polynomials):
\begin{eqnarray}
\psi_m(x) = \psi(x_m) + \psi'(x_m) (x-x_m) + \frac{1}{2} \psi''(x_m) (x-x_m)^2 + ...
\label{MagnusTaylor}
\end{eqnarray}
allowing to represent the Magnus expansion in powers of $\Delta x_m$ leading to a systematic algorithm for building numerical schemes of a certain order.
The trace-vanishing feature of the matrix $\widehat{\mathbf{Q}}$ allows to represent its exponential as follows:
\begin{eqnarray}
e^{\widehat{\mathbf{\Omega}} (\tau_m)} =
\begin{pmatrix}
\cosh k_m - \frac{i \varkappa_m}{k_m} \sinh k_m & \frac{\chi_m}{k_m} \sinh k_m \\ \frac{- \chi^*_m}{k_m} \sinh k_m & \cosh k_m + \frac{i \varkappa_m}{k_m} \sinh k_m
\end{pmatrix},
\label{eqMexp1}
\end{eqnarray}
%
%where $k_m^2 = (\sigma |q_m|^2 - \zeta^2) \Delta \tau_m^2$, $\zeta_m = \zeta \Delta \tau_m$ and $\chi_m = q_m \Delta \tau_m$.
%
where the coefficients are expressed for the sixth-order scheme:
%
%\begin{eqnarray}
%
%\label{km2} && k_m^2 = (\sigma |q_m|^2 - \zeta^2) \Delta \tau_m^2 + \sigma (q_m^* q^{\prime\prime}_m + q_m^{\prime\prime *} q_m) \Delta \tau_m^4 / 24, \;\;\;\;\;\; \\
%
%\label{zm} && i \zeta_m = i \zeta \Delta \tau_m + \sigma (q_m^* q'_m - q_m q_m^{\prime *}) \Delta \tau_m^3 / 12, \\
%
%\label{chim} && \chi_m = \underbrace{q_m \Delta \tau_m}_\text{2nd} + \underbrace{(q''_m + 4 i q'_m \zeta) \Delta \tau_m^3 / 24}_\text{4th order},
%
%
%\end{eqnarray}
%
%
\begin{eqnarray}
%\varkappa
            && k_m^2 = \textcolor{red}{(- |\psi_m|^2 - \lambda^2) \Delta x_m^2}     \textcolor{OliveGreen}{- (\psi_m^* \psi^{\prime\prime}_m + \psi_m^{\prime\prime *} \psi_m) \Delta x_m^4 / 24}     \textcolor{blue}{+     8 ( \psi_m^* \psi_m' - \psi_m \psi_m'^* )^2   \Delta x_m^6 / 5760 +} \nonumber\\
\label{km2} && \textcolor{blue}{+ \{ 3 (\psi_m^* \psi_m'''' + \psi_m \psi_m''''^*) + 10 |\psi_m''|^2 + 16 i \lambda (\psi_m' \psi_m''^* - \psi_m'^* \psi_m'')  - 32 |\psi_m'|^2 \lambda^2\} \Delta x_m^6 / 5760}, \\
            && i \varkappa_m = \textcolor{red}{i \lambda \Delta x_m} \textcolor{OliveGreen}{+ (\psi_m^* \psi'_m - \psi_m \psi_m^{\prime *}) \Delta x_m^3 / 12 }      \textcolor{blue}{+     8 |\psi_m|^2 (\psi_m^* \psi_m' - \psi_m \psi_m'^*) \Delta x_m^5 / 1440      + \{ 3 ( \psi_m^* \psi_m''' - \psi_m \psi_m'''^* )  -} \nonumber\\
\label{zm}  && \textcolor{blue}{- 3 (\psi_m'^* \psi_m'' - \psi_m' \psi_m''^*) - 4 i \lambda (\psi_m^* \psi_m'' + \psi_m \psi_m''^* - 6 |\psi_m'|^2 ) - 8 \lambda^2 (\psi^* \psi' - \psi \psi'^*) \} \Delta x_m^5 / 1440}, \\
            && \chi_m = \textcolor{red}{\psi_m \Delta x_m} \textcolor{OliveGreen}{+ (\psi''_m + 4 i \psi'_m \lambda) \Delta x_m^3 / 24  }   \textcolor{blue}{ +     \{  3 \psi_m'''' + 24 i \psi_m''' \lambda - 32 \psi_m'' \lambda^2 + 64 i \psi_m' \lambda^3  \} \Delta x_m^5 / 5760 +} \nonumber\\
\label{chim} && \textcolor{blue}{+ \{ 48 ( \psi_m^* \psi_m'^2 - \psi_m |\psi_m'|^2 ) - 16 (|\psi_m|^2 \psi'' - \psi^2 \psi''^*) + 64 i |\psi_m|^2 \psi_m' \lambda \} \Delta x_m^5 / 5760},
\end{eqnarray}
with $\psi'_m = \psi'(x_m)$, $\psi''_m = \psi''(x_m)$, $\psi'''_m = \psi'''(x_m)$ and $\psi''''_m = \psi''''(x_m)$ and the color code corresponding to the scheme order as in Fig. \ref{fig:2}.
First terms (in red) in Eqs. (\ref{km2})-(\ref{chim}) correspond to the Boffetta--Osborne second-order scheme.
To obtain this low-order scheme one has to retain only the first term in the expansion (\ref{MagnusOmega}) for $\widehat{\mathbf{\Omega}}$ and in the expansion (\ref{MagnusTaylor}) for $\psi(x)$ inside the bin.
Increasing the number of terms in (\ref{MagnusOmega}) and (\ref{MagnusTaylor}) we derive fourth- and sixth-order schemes represented by green and blue color in Eqs. (\ref{km2})-(\ref{chim}), respectively.
The matrix with $\lambda$-derivatives $\widehat{\mathbf{U}}'(x_m) = \partial_\lambda \widehat{\mathbf{U}}$ which is connected to $\bold{\widehat{\Sigma}'}$ has the following elements $U'_{ml}$, where $m,l=1,2,3,4$:
\begin{eqnarray}
&& U'_{11} = \partial_\lambda \Big(\cosh k_m - \frac{i \varkappa_m}{k_m} \sinh k_m \Big) = \Big( k_m' - \frac{i \varkappa_m'}{k_m} + \frac{i \varkappa_m k_m'}{k_m^2} \Big) \sinh k_m - \frac{i \varkappa_m k_m'}{k_m} \cosh k_m, \\
&& U'_{12} = \partial_\lambda \Big( \frac{\chi_m}{k_m} \sinh k_m \Big) = \Big( \frac{\chi_m'}{k_m} - \frac{\chi_m k_m'}{k_m^2} \Big) \sinh k_m + \frac{\chi_m k_m'}{k_m} \cosh k_m, \\
&& U'_{21} = \partial_\lambda \Big( -\frac{\chi^*_m}{k_m} \sinh k_m \Big) = - \Big( \frac{\chi_m'^*}{k_m} - \frac{\chi_m^* k_m'}{k_m^2} \Big) \sinh k_m - \frac{\chi_m^* k_m'}{k_m} \cosh k_m, \\
&& U'_{22} = \partial_\lambda \Big( \cosh k_m + \frac{i \varkappa_m}{k_m} \sinh k_m \Big) = \Big( k_m' + \frac{i \varkappa_m'}{k_m} - \frac{i \varkappa_m k_m'}{k_m^2} \Big) \sinh k_m + \frac{i \varkappa_m k_m'}{k_m} \cosh k_m,
\label{eq4orUzeta}
\end{eqnarray}
%
%
%The sixth-order scheme is obtained in a similar fashion considering $\Omega_i$ in Eqs. (\ref{Om1})-(\ref{Om4}) up to $\Omega_4$ and keeping terms proportional to $\Delta \tau_m^6$ and $\Delta \tau_m^5$ in Eqs. (\ref{km2}), (\ref{zm}) and (\ref{chim}), respectively.
%
%
where $k_m' = \partial_\lambda k_m$, $\chi_m' = \partial_\lambda \chi_m$, $\varkappa_m' = \partial_\lambda \varkappa_m$ can be obtained from (\ref{km2})-(\ref{chim}):
%
%where $k_m' = \partial_\lambda k_m$, $\chi_m' = \partial_\lambda \chi_m$ и $\varkappa_m' = \partial_\lambda \varkappa_m$ can be obtained from (\ref{km2})-(\ref{chim}):
%
\begin{eqnarray}
\label{km2dzeta}  && k_m' = \frac{1}{2 k_m} \Big( -2 \lambda \Delta x_m^2 + \{ 16 i (\psi_m' \psi_m''^* - \psi_m'^* \psi_m'') - 64 |\psi_m'|^2 \lambda \} \frac{\Delta x_m^6}{5760} \Big), \\
\label{zmdzeta}   && \varkappa_m' = \Delta x_m - \Big(  4 \{ \psi_m^* \psi_m'' + \psi_m \psi_m''^* - 6 |\psi_m'|^2 \} - 16 i \lambda \{ ^* \psi' - \psi \psi'^* \}  \Big) \frac{\Delta x_m^5}{1440}, \\
\label{chimdzeta} && \chi_m' = 4 i \psi_m' \frac{\Delta x_m^3}{24} + \Big( 24 i \psi_m'''  - 64 \lambda \psi_m'' + 192 i \lambda^2 \psi_m' + 64 i |\psi_m|^2 \psi_m' \Big) \frac{\Delta x_m^5}{5760}.
\end{eqnarray}

\end{document}